\documentclass[conference]{IEEEtran}
\usepackage{amsmath,amssymb,graphicx,mathtools}
\usepackage{algorithm,algorithmic,multicol,tikz,cite,array,booktabs,url,flushend}

\usepackage{circuitikz}
\usetikzlibrary{angles,quotes}
\usetikzlibrary{patterns}
\usetikzlibrary{shadows.blur}
\usetikzlibrary{positioning}
\tikzset{font={\fontsize{12pt}{12}\selectfont}}

\title{A Hardware Architecture for Reconfigurable Intelligent Surfaces with Minimal Active Elements for Explicit Channel Estimation}
\author{
\IEEEauthorblockN{George~C.~Alexandropoulos$^1$ and Evangelos Vlachos$^2$
 }
\IEEEauthorblockA{
$^1$Department of Informatics and Telecommunications, National and Kapodistrian University of Athens,\\
Panepistimiopolis Ilissia, 15784 Athens, Greece\\
$^2$Institute for Digital Communications, University of Edinburgh, EH9 3JL Edinburgh, UK\\
emails: alexandg@di.uoa.gr, e.vlachos@ed.ac.uk
}}

\begin{document}

\maketitle

\begin{abstract}
Intelligent surfaces comprising of cost effective, nearly passive, and reconfigurable unit elements are lately gaining increasing interest due to their potential in enabling fully programmable wireless environments. They are envisioned to offer environmental intelligence for diverse communication objectives, when coated on various objects of the deployment area of interest. To achieve this overarching goal, the channels where the Reconfigurable Intelligent Surfaces (RISs) are involved need to be in principle estimated. However, this is a challenging task with the currently available hardware RIS architectures requiring lengthy training periods among the network nodes utilizing RIS-assisted wireless communication. In this paper, we present a novel RIS architecture comprising of any number of passive reflecting elements, a simple controller for their adjustable configuration, and a single Radio Frequency (RF) chain for baseband measurements. Capitalizing on this architecture and assuming sparse wireless channels in the beamspace domain, we present an alternating optimization approach for explicit estimation of the channel gains at the RIS elements attached to the single RF chain. Representative simulation results demonstrate the channel estimation accuracy and achievable end-to-end performance for various training lengths and numbers of reflecting unit elements. 

%Capitalizing on this architecture, we present an algorithm for RIS tuning that includes a channel estimation stage for the surface's portion attached to the RF chains. The role of the size of this portion in the achievable RIS-assisted performance is investigated via simulations over wireless fading channels with a dominant line-of-sight component.
\end{abstract}


\section{Introduction}
The increasingly demanding objectives for beyond fifth Generation (5G) communications have spurred recent research activities
on novel wireless hardware architectures \cite{Alkhateeb_JSTSP_2014, alexandg_ESPARs, sha_hu}. Among them belong the Reconfigurable Intelligent Surfaces (RISs) \cite{Kaina_metasurfaces_2014, Chen2016, Yang_metasurfaces_2016, Foo_LC}, which are artificial planar structures with integrated electronic circuits that can be programmed to manipulate an incoming ElectroMagnetic (EM) field in a wide variety of functionalities. Incorporating RISs in wireless networks has been recently envisioned as a revolutionary means to transform any passive wireless communication environment to an active reconfigurable one \cite{Liaskos_Visionary_2018, Marco_Visionary_2019, Liaskos_TN_2019}, offering environmental intelligence for diverse communication objectives.

%(ranging from perfect and controllable absorption, beam and wavefront shaping to polarization control, broadband pulse delay, and harmonic generation)
Differently from conventional relaying systems \cite{George_RIS_TWC2019, Emil_Relay_2019}, recent RIS designs are mainly based on metamaterials and are comprised of periodically aligned subwavelength elements, termed as unit cells, which are capable of offering overall control over the metasurface's EM behavior \cite{Liaskos_Visionary_2018, Marco_Visionary_2019, LISA_RIS_2019}. Considering RIS as a means to assist data communication between a Base Station (BSs) and User Equipments (UEs), \cite{George_RIS_TWC2019, Wu_RIS_TWC2019} presented design algorithms for the phase shifting values of the RIS unit cells under the assumption of perfect availability of the wireless channels where RIS is involved. However, estimating those channels and then sharing this information is a challenging task with the currently available metasurface architectures, which consist of nearly passive elements. Very recently, \cite{Deepak_RIS_ICASSP2019, He_RIS_CE_2019, Zhang_RIS_CE_2019} proposed algorithmic approaches with relevant communication protocols and RIS training configurations to estimate the concatenated channel among the BS, RIS, and UE at either the BS or UE sides, which require, however, large training periods. Alternatively, \cite{George_RIS_DNN_SPAWC2019, Alkhateeb_RIS_CS_2019} demonstrated that dedicated deep neural networks can be trained during appropriately designed offline phases to provide efficient online RIS configuration for specific synthetic indoor and outdoor environments. Specifically, in \cite{Alkhateeb_RIS_CS_2019}, the authors proposed a RIS architecture comprising of passive reflecting elements and multiple active elements that are connected to baseband for partial channel estimation.

In this paper, we present a novel RIS architecture comprising of passive unit elements and a single active Radio Frequency (RF) chain for baseband reception, which enables explicit channel estimation at the RIS side. The proposed architecture is inspired by the recently proposed extended analog combiner in \cite{Vlachos2019WidebandSampling} enabling matrix-completion-based channel estimation \cite{Vlachos_SPL2018} with relatively short training requirements. 

\section{System and Channel Models}
Consider a single-antenna BS wishing to communicate in the downlink direction with a single-antenna UE. We assume that there is no direct link between BS and UE due to blockages, and that a planar RIS equipped with $N\triangleq N_{\rm v}N_{\rm h}$ unit elements ($N_{\rm h}$ in the horizontal and $N_{\rm v}$ in the vertical orientations) is deployed to enable reliable and high data rate wireless communication between BS and UE. The proposed solutions and results in this paper can be extended to multi-antenna BS and UE, uplink communication, and to the case where the direct link exists. The complex-valued baseband received signal at UE can be expressed as \cite[eq. (1)]{George_RIS_TWC2019}
\begin{equation}\label{system_model}
  y = \mathbf{h}_{2}\mathbf{\Phi}\mathbf{h}_{1}s+w = \left(\mathbf{h}_{2}\circ\mathbf{h}_{1}^T\right)\boldsymbol{\phi}s+w,
\end{equation}
where $\mathbf{h}_{1}\in\mathbb{C}^{N\times 1}$ denotes the channel vector between RIS and BS and $\mathbf{h}_{2}\in\mathbb{C}^{1\times N}$ is the channel vector between UE and RIS with operand %$(\cdot)^{\rm T}$ representing vector transposition and 
$\circ$ representing the Hadamard product. In addition, $\mathbf{\Phi}\triangleq\mathrm{diag}\{\boldsymbol{\phi}\}\in\mathbb{C}^{N\times N}$, with $\boldsymbol{\phi}\in\mathbb{C}^{N\times 1}$, is a diagonal matrix accounting for the effective phase shifts applied by the RIS unit elements, where $[\boldsymbol{\phi}]_n=e^{j\theta_n}$ $\forall$$n=1,2,\ldots,N$ with $j\triangleq\sqrt{-1}$ being the imaginary unit. The symbol $s$ in \eqref{system_model} denotes the unit power complex-valued information symbol chosen from a discrete constellation set, and $w\sim\mathcal{CN}(0,\sigma^2)$ models the zero mean complex Additive White Gaussian Noise (AWGN) with variance $\sigma^2$. In this paper, we consider finite resolution phase shifting values for the RIS unit cells, and particularly, each $[\boldsymbol{\phi}]_n$ is obtained as
\begin{equation}\label{phase shifting_model}
[\boldsymbol{\phi}]_n \in \mathcal{F} \triangleq \left\{e^{j2^{1-b}\pi m}\right\}_{m=0}^{2^{b}-1},
\end{equation}
where $\mathcal{F}$ represents each cell's reflection feasible set and $b$ is the phase resolution in number of bits. Clearly, the different number of phase shifting values per RIS unit element is $2^b$. 

Each of the wireless channels $\mathbf{h}_1$ and $\mathbf{h}_2$ is assumed to be comprised of $N_p$ propagation paths. Let $\alpha_{{\rm R},k}$ and $\alpha_{{\rm T},k}$ represent the gains of the $k$-th paths ($k=1,2,\ldots,N_p$) in the BS to RIS and the RIS to UE channels, respectively, that are drawn from the zero mean complex Gaussian noise distribution with variances $(2{\rm PL}_1)^{-1}$ and $(2{\rm PL}_2)^{-1}$. Notations ${\rm PL}_1$ and ${\rm PL}_2$ refer to the path losses in the BS to RIS and RIS to UE channels, respectively. In general far field cases, each path loss scales with the inverse square of the distance between the involved terminals. Moreover, we denote by $\mathbf{a}_{\rm R}(\phi_{\rm R}^{(k)},\theta_{\rm R}^{(k)})\in\mathbb{C}^{N\times1}$ and $\mathbf{a}_{\rm T}(\phi_{\rm T}^{(\ell)},\theta_{\rm T}^{(\ell)})\in\mathbb{C}^{N\times1}$ the response vectors for the BS to RIS and RIS to UE channels, respectively, where $\phi_{\rm R}^{(k)}$, $\theta_{\rm R}^{(k)}$ and $\phi_{\rm T}^{(k)}$, $\theta_{\rm T}^{(k)}$ represent the physical elevation and azimuth angles of arrival and departure, respectively. Using the latter definitions, the considered models for the wireless channels $\mathbf{h}_1$ and $\mathbf{h}_2$ are expressed as \cite{Alkhateeb_JSTSP_2014}
\begin{subequations}
\begin{equation}
\mathbf{h}_1 \triangleq \sum_{k=1}^{N_p} \alpha_{{\rm R},k} \mathbf{a}_{\rm R}(\phi_{\rm R}^{(k)}, \theta_{\rm R}^{(k)}),
\end{equation}
\begin{equation}
\mathbf{h}_2 \triangleq \sum_{k=1}^{N_p} \alpha_{{\rm T},k} \mathbf{a}_{\rm T}^H(\phi_{\rm T}^{(k)}, \theta_{\rm T}^{(k)}).
\end{equation}
\end{subequations}
We further adopt the beamspace model of \cite{Sayeed_TAP_2013} to alternatively express the channels as $\mathbf{h}_1 = \mathbf{D}_{\rm R}\mathbf{z}_1$ and $\mathbf{h}_2 = \mathbf{z}_2^H\mathbf{D}_{\rm T}^H$, where $\mathbf{D}_{\rm R}\in\mathbb{C}^{N\times N}$ and $\mathbf{D}_{\rm T}\in\mathbb{C}^{N\times N}$ are unitary matrices based on the discrete Fourier transform, and $\mathbf{z}_1\in\mathbb{C}^{N\times1}$ and $\mathbf{z}_2\in\mathbb{C}^{N\times1}$ respectively contain only few virtual channel gains with high amplitude (i.e., they are sparse vectors). The latter is particularly true for large $N$ values (e.g., $N=102$ in \cite{Kaina_metasurfaces_2014}'s RIS design) and/or millimeter wave channels. We note that the consideration of possible mutual coupling among the RIS elements in the channel model is left for future work. 
\begin{figure}
    \centering
    \includegraphics[scale=0.5]{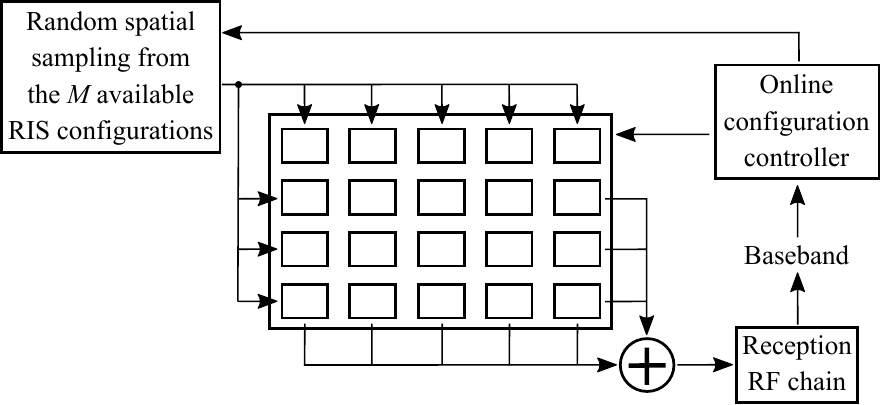}
    \caption{Block diagram of the proposed RIS hardware architecture including a single active reception RF chain for explicit channel estimation at the RIS side.}
\label{fig:RIS_architecture}
\end{figure}

\section{Proposed RIS Architecture}
Current RIS designs with nearly passive elements \cite{Kaina_metasurfaces_2014, Chen2016, Yang_metasurfaces_2016, Foo_LC} are based on RF micro electromechanical systems or metamaterials that enable real-time control of the elements' reflection coefficients. However, explicit estimation at RIS side of the channel gains at its elements is not feasible. In this paper, we propose to connect the outputs of the RIS unit elements to a single reception RF chain, as shown in Fig$.$~\ref{fig:RIS_architecture}. This single RF chain enabling baseband channel estimation via pilot signals consists of a low noise amplifier, a mixer downconverting the signal from RF to baseband, and an analog to digital converter. Each impinging EM signal at the RIS unit cells that carries a training symbol, is received in the RF domain with one of the $M\leq\mathcal{S}^N_\mathcal{F}$ (with $\mathcal{S}_\mathcal{F}$ denoting the cardinality of set $\mathcal{F}$) available RIS configurations in the random sampling unit; this configuration is selected in a manner resulting in random spatial sampling. Random analog combining has been recently proposed in \cite{Vlachos2019WidebandSampling} for millimeter wave channel estimation. The $M$ available RIS configurations at the random sampler are represented by $\mathbf{W} \in \mathcal{W}^{M \times N}$ with $\mathcal{W}$ denoting a finite set of complex-valued matrices with unit amplitude and quantized phase elements as in \eqref{phase shifting_model}. Each column of $\mathbf{W}$ refers to an analog receiver that can spatially process the impinging to the RIS elements  EM signal before feeding it to the sole RF chain through the adders, as illustrated in Fig$.$~\ref{fig:RIS_architecture}.    

The proposed RIS architecture enables accurate estimation of the RIS involved channels $\mathbf{h}_1$ and $\mathbf{h}_2$, as will described in the sequel. As shown in Fig$.$~\ref{fig:RIS_architecture}, it also includes a dedicated control unit for implementing the channel estimation algorithm, calculating the RIS configuration tuning, and sharing the phase shifting values to all of its $N$ unit elements. We note that, differently from \cite{Alkhateeb_RIS_CS_2019}'s RIS architecture that includes $N$ RF chains for $N$ active channel sensors, our propposed RIS hardware architecture has only one RF chain.      

\section{RIS Channel Estimation and Tuning}
In this section, we capitalize on the proposed RIS architecture in Fig$.$~\ref{fig:RIS_architecture} and present a multi-objective optimization framework for explicit channel estimation at the RIS side.  

\subsection{Proposed Channel Estimation Formulation}
During dedicated training slots per channel access, the BS sends $T$ training symbols to RIS, which are assumed to be known at both sides. At each of those slots, each training signal is received at the $N$ RIS unit elements. In particular, for the reception of each of the latter signals, the phase shifting values at the RIS elements are configured with one of the $M$ available configurations that is randomly chosen from the random spatial sampling unit. Then, as shown in Fig$.$~\ref{fig:RIS_architecture}, the output signals from the RIS elements are all summed together and fed to single RF chain. We denote by $\mathbf{y}_t \in\mathbb{C}^{N\times 1}$ the $t$-th received training signal with $t=1,2,\ldots,T$ at the $N$ RIS unit elements, which can be mathematically expressed as 
\begin{equation}\label{eq:y_t}
\mathbf{y}_t\triangleq\mathbf{h}_1q_t+\mathbf{n}_t,
\end{equation}
with $q_t$ being the $t$-th training symbol and $\mathbf{n}_t \in \mathbb{C}^{N\times 1}$ is the AWGN matrix having independent and identically distributed elements as $\mathcal{CN}(0,{\rm SNR}^{-1})$, where the transmit Signal to Noise Ratio (SNR) is defined as the ratio of transmit power over the AWGN power. 

Given the matrix $\mathbf{W}$ including the $M$ available configurations for the RIS elements, the output of the random spatial sampling unit for the $t$-th symbol can be represented by $\mathbf{r}_t\in \mathbb{C}^{M\times 1}$ defined as 
\begin{equation}
\mathbf{r}_t\triangleq\boldsymbol{\omega}_t\circ\mathbf{W}\mathbf{y}_t,
\end{equation}
where the $M\times 1$ vector $\boldsymbol{\omega}_t$ is composed of one unity element and $M-1$ zeros. The position of the unity is chosen randomly in a uniform fashion over the set $\{1,2,\ldots,M\}$. This is equivalent to randomly choosing a $\mathbf{W}$'s row (i.e., one RIS configuration) and then feeding the outputs of the RIS elements to the single RF chain. Particularly, the sole non-zero element of $\mathbf{r}_t$ represents the input to the RF chain. We next use the notation $\mathbf{\Omega} \triangleq [\boldsymbol{\omega}_1\,\boldsymbol{\omega}_2\cdots\, \boldsymbol{\omega}_T]$ for the $M\times T$ matrix having as columns the selections for the inputs to the single RF chain for all RIS received training signals. Based on expression \eqref{eq:y_t}, the $T$ inputs to the RIS RF chain referring to the $T$ received training signals can be compactly expressed with $\mathbf{R}_{\Omega}\in\mathbb{C}^{M\times T}$, which is given by
\begin{equation}
\mathbf{R}_{\Omega} \triangleq \mathbf{\Omega} \circ (\mathbf{W} \mathbf{Y}),
\end{equation}
where $\mathbf{Y}\triangleq [\mathbf{y}_1\,\mathbf{y}_2\,\cdots\,\mathbf{y}_T]\in\mathbb{C}^{N\times T}$. To increase the baseband measurements for $\mathbf{h}_1$ estimation at RIS, one may trivially increase $T$ or use more than one RIS configurations for receiving each training signal (i.e., each of the training symbols spans more than one training slots).

Capitalizing on the beamspace representation $\mathbf{h}_1 = \mathbf{D}_{\rm R}\mathbf{z}_1$ and the sparse structure of $\mathbf{z}_1$, the noiseless received training signal matrix $\tilde{\mathbf{Y}}\triangleq  \mathbf{W}\mathbf{h}_1\mathbf{q}\in\mathbb{C}^{M\times T}$ (where $\mathbf{q}\triangleq [q_1\,q_2\,\cdots\,q_T]$ includes the $T$ available training symbols) has a low rank in the beamspace domain. Using this property, we formulate the following multi-objective optimization problem for the $\mathbf{h}_1$ estimation at the RIS side:  
\begin{align}
\textrm{min}_{\tilde{\mathbf{Y}}, \mathbf{z}_1} \,\,\, &\tau_R \Vert \tilde{\mathbf{Y}} \Vert_* + \tau_Z \Vert \mathbf{z}_1 \Vert_1 \nonumber + \frac{1}{2} \Vert \mathbf{R}_{\Omega} - (\mathbf{\Omega}\circ \tilde{\mathbf{Y}}\Vert_F^2 \nonumber \\
\textrm{s.t.} \,\,\, &\mathbf{Y} = \mathbf{W}\mathbf{D}_{\rm R} \mathbf{z}_1\mathbf{q}  + \mathbf{N},  \label{eq:op}
\end{align}
where $\mathbf{N}\triangleq [\mathbf{n}_1\,\mathbf{n}_2\,\cdots\,\mathbf{n}_T]\in\mathbb{C}^{N\times T}$ the nuclear norm $\Vert \tilde{\mathbf{Y}} \Vert_* $ imposes the low rank property to $\tilde{\mathbf{Y}}$, while the $\ell_1$-norm $\Vert \mathbf{z}_1 \Vert_1$ enforces the sparse structure of $\mathbf{z}_1$. The weighting factors $\tau_R, \tau_Z>0$ depend in general on the number $N_p$ of the distinct propagation paths in the BS to RIS channel $\mathbf{h}_1$. We finally note that $\mathbf{h}_2$'s estimation follows a similar procedure to $\mathbf{h}_1$ estimation using training symbols sent from the UE to RIS. If these symbols are orthogonal to the training symbols sent from the BS to RIS, the estimation of $\mathbf{h}_1$ and $\mathbf{h}_2$ can take place in parallel.

\subsection{An ADMM-Based Algorithm}
The optimization in \eqref{eq:op} includes a highly coupled problem, thus, to be efficiently solved, its cost function has to be decomposed into multiple simpler subproblems. To this end, we first introduce the two auxiliary matrix variables $\mathbf{X}, \mathbf{C} \in\mathbb{C}^{M \times T}$, which are deployed in re-expressing \eqref{eq:op} as follows:
\begin{align}\label{eq:mc_opt_side_information_splitted}
\min_{\tilde{\mathbf{Y}}, \mathbf{z}_1, \mathbf{X}, \mathbf{C}}& \tau_R \Vert \tilde{\mathbf{Y}} \Vert_* + \tau_Z \Vert \mathbf{z}_1 \Vert_1 + \frac{1}{2} \Vert \mathbf{C} \Vert_F^2 \nonumber \\ & + \frac{1}{2} \Vert  \mathbf{R}_{\Omega} - (\mathbf{\Omega}\circ \tilde{\mathbf{Y}}) \Vert_F^2 \nonumber \\
\textrm{s.t.}& \, \tilde{\mathbf{Y}} = \mathbf{X} \,\text{ and } \, \mathbf{C} = \mathbf{X} - \mathbf{W} \mathbf{D}_{\rm R} \mathbf{z}_1\mathbf{q}.
\end{align}
The Lagrangian function of the latter problem is given by
\begin{align}\label{eq:L}
& \mathcal{L}\left(\tilde{\mathbf{Y}}, \mathbf{z}_1, \mathbf{X}, \mathbf{C}, \mathbf{V}^{(1)}, \mathbf{V}^{(2)}\right) \triangleq \tau_R \Vert \tilde{\mathbf{Y}} \Vert_* + \tau_Z \Vert \mathbf{z}_1 \Vert_1 \nonumber \\& 
+ \frac{1}{2} \Vert \mathbf{C} \Vert_F^2 + \frac{1}{2} \Vert \mathbf{R}_{\Omega} - (\mathbf{\Omega} \circ \mathbf{X}) \Vert_F^2 + \textrm{tr}\{\mathbf{V}_1^H (\mathbf{Y} - \mathbf{X})\} 
\nonumber \\&
 + \frac{\gamma}{2} \Vert \mathbf{Y} - \mathbf{X} \Vert_F^2 + \textrm{tr}\{\mathbf{V}_2^H (\mathbf{C} - \mathbf{X} + \mathbf{W}\mathbf{D}_{\rm R}\mathbf{z}_1\mathbf{q})\} \nonumber \\ & + \frac{\gamma}{2} \Vert \mathbf{C} - \mathbf{X} + \mathbf{W} \mathbf{D}_{\rm R}\mathbf{z}_1\mathbf{q} \Vert_F^2,
\end{align}
where $\mathbf{V}^{(1)}, \mathbf{V}^{(2)} \in \mathbb{C}^{M \times T}$ are the Lagrange multipliers (i.e., the dual variables) adding the constraints of \eqref{eq:mc_opt_side_information_splitted} to the cost function, and $\gamma\in(0,1)$ denotes the ADMM's stepsize. The basic steps of \eqref{eq:mc_opt_side_information_splitted}'s solution are summarized in Algorithm~\ref{algorithm:proposed_algorithm}. We omit further details for this algorithm due to space limitations, however, a similar procedure to \cite{Vlachos2019WidebandSampling} has been be followed. For the initialization at $n=0$: $\tilde{\mathbf{Y}}_0 = \mathbf{X}_0 = \mathbf{C}_0 = \mathbf{V}_0^{(1)} = \mathbf{V}_0^{(2)} = \mathbf{0}_{M \times T}$. At line 8, the symbol $\dagger$ represents the pseudo-inverse operation, while SVT stands for the Singular Value Thresholding operator \cite[eq. (36)]{Vlachos2019WidebandSampling}.
\begin{algorithm}[!t]
	\caption{Proposed Single-RF RIS Channel Estimation}
	\begin{algorithmic}[1]
		\REQUIRE $\mathbf{\Omega}$, $\mathbf{D}_{{\rm R}}$, $\mathbf{W}$, $\mathbf{q}$, $\tau_R$, $\tau_Z$, $\gamma$, and $I_{\rm max}$.
		\ENSURE $\mathbf{z}_1$
		
		\FOR {$n=1,2,\ldots,I_{\rm max}$}
		\STATE \COMMENT{\textbf{Minimization of $\mathcal{L}$ in \eqref{eq:L} over $\tilde{\mathbf{Y}}_n$}}
        \STATE $\tilde{\mathbf{Y}}_n = \textrm{SVT}\{\mathbf{X}_{n-1} - \gamma^{-1} \mathbf{V}^{(1)}_{n-1}\}$.

        \STATE \COMMENT{\textbf{Minimization of $\mathcal{L}$ in \eqref{eq:L} over $\mathbf{X}_n$}}
		\STATE Solve the following linear system of equations:
		\begin{align*}
		& \left( \sum_{j=1}^{M} \textrm{diag}([\mathbf{\Omega}]_j)^T \otimes \mathbf{E}_{jj}  + 2 \gamma \mathbf{I}_{T M} \right) \mathbf{x}_n =  \\
		& \text{vec}\{\gamma \tilde{\mathbf{Y}}_{n-1} \!+\! \mathbf{R}_{\Omega} \!+\! \mathbf{V}^{(1)}_{n-1}+\mathbf{V}^{(2)}_{n-1} \!+\! \gamma \mathbf{C}_{n-1}\} \\ & + \gamma (\mathbf{q}^T \otimes \mathbf{W} \mathbf{D}_{\rm R}) \mathbf{z}_1
		\end{align*}
		with $\mathbf{E}_{jj}$ obtained from the $M\times M$ all-zero matrix after inserting a unity value at its $(j,j)$-th position.
		\STATE Reshape vector $\mathbf{x}_n$ to matrix $\mathbf{X}_n$: $\mathbf{X}_n=\mathrm{unvec}\{\mathbf{x}_n\}$.
        
        \STATE \COMMENT{\textbf{Minimization of $\mathcal{L}$ in \eqref{eq:L} over $\mathbf{z}_1$}}
        \STATE Set $\boldsymbol{\xi} = (\mathbf{q}^T \otimes \mathbf{W} \mathbf{D}_{\rm R})^{\dagger} \text{vec}\{\mathbf{X}_n-\mathbf{C}_{n-1}-\gamma^{-1} \mathbf{V}^{(2)}_{n-1}\}$.
        \STATE Apply the following soft-thresholding operator:
        \begin{align*}\mathbf{z}_1=&\mathrm{sign}\{\mathtt{Re}(\boldsymbol{\xi})\} \circ \max \big\{\vert \mathtt{Re}(\boldsymbol{\xi}) \vert - \tau_Z\gamma^{-1}, 0 \big\} \nonumber \\ &+ \mathrm{sign}\{\mathtt{Im}(\boldsymbol{\xi})\} \circ \max \big\{\vert \mathtt{Im}(\boldsymbol{\xi}) \vert - \tau_Z\gamma^{-1},0 \big\}.
        \end{align*}
        \STATE \COMMENT{\textbf{Minimization of $\mathcal{L}$ in \eqref{eq:L} over $\mathbf{C}_n$}}
        \STATE Set $\mathbf{C}_n = \frac{\gamma}{1+\gamma} \left(\mathbf{X}_n -  \mathbf{W} \mathbf{D}_{\rm R} \mathbf{z}_1\mathbf{q} - \gamma^{-1}\mathbf{V}^{(2)}_{n-1} \right)$.
        
        \STATE \COMMENT{\textbf{Update the dual variables}}
        \STATE $\mathbf{V}^{(1)}_n\!\!=\!\mathbf{V}^{(1)}_{n-1} + \gamma \big( \mathbf{X}_n\!-\!\tilde{\mathbf{Y}}_n\big)$.
        \STATE $\mathbf{V}^{(2)}_n\!\!=\!\mathbf{V}^{(2)}_{n-1} \!+\! \gamma \big( \mathbf{C}_n \!-\! \mathbf{X}_n \!+\! \mathbf{W} \mathbf{D}_{\rm R} \mathbf{z}_1\mathbf{q} \big)$.
        \ENDFOR
	\end{algorithmic}
	\label{algorithm:proposed_algorithm}
\end{algorithm}

\subsection{Online Tuning of the RIS Unit Elements}\label{sec:tuning}
Given the estimates $\hat{\mathbf{h}}_1$ and $\hat{\mathbf{h}}_2$ for the channels $\mathbf{h}_1$ and $\mathbf{h}_2$, respectively, at the RIS controller using Algorithm~1, the optimum configuration for the reflection coefficients of its $N$ elements maximizing the achievable end-to-end rate is obtained from the following optimization problem ($n=1,2,\ldots,N$):
\begin{equation}\label{eq:optimum_phase_shift}
\max_{\boldsymbol{\phi}}\log_2\left(1+\left|\left(\hat{\mathbf{h}}_{2}\circ\hat{\mathbf{h}}_{1}^{\rm T}\right)\boldsymbol{\phi}\right|^2\right)\,\,{\rm s.t.}\,\,[\boldsymbol{\phi}]_n \in \mathcal{F}\,\,\forall n.%n=1,2,\ldots,N
\end{equation}
The RIS controller may solve \eqref{eq:optimum_phase_shift} via exhaustive search for reasonable numbers of the available RIS configurations (i.e., affordable $M$ or $\mathcal{S}^N_\mathcal{F}$). Alternatively, each $[\boldsymbol{\phi}]_n$ can be set as the closest feasible value in $\mathcal{F}$ to its optimum value $e^{j\theta_n}$, where the phase shift $\theta_n$ is given by  
\begin{equation}\label{eq:optimum_phase_shift_1}
\theta_n=-\arg\left([\hat{\mathbf{h}}_1]_n[\hat{\mathbf{h}}_2]_n\right).
\end{equation}

\section{Simulation Results} 
\begin{figure}
    \centering
		\includegraphics[scale=0.47]{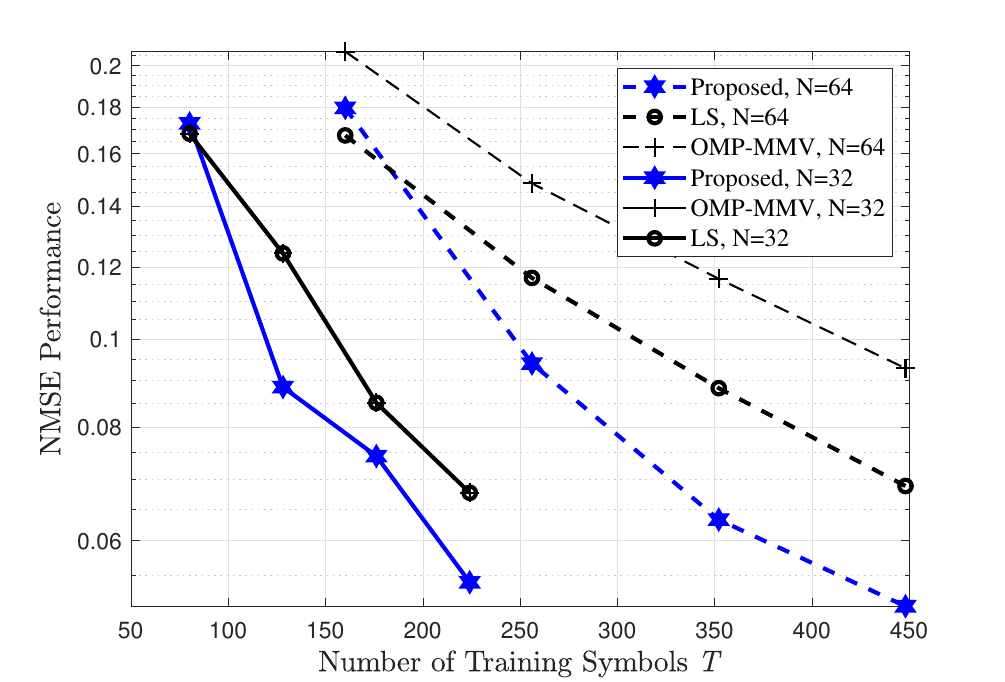}
    \caption{NMSE vs the training symbols' number $T$ for $b=4$, different values for the $N$ RIS elements, and ${\rm SNR}=5$dB.}
    %with AWGN variance $\sigma_n^2=0.3162$.}
    \label{fig:nmseVSframes_ris}
\end{figure}

We commence with the performance evaluation of the proposed channel estimation technique focusing on the BS to RIS channel. In Fig$.$~\ref{fig:nmseVSframes_ris}, we have considered a RIS with the proposed single-RF architecture for $N=32$ and $64$ unit elements. We have also set the phase resolution $b=4$ and the available RIS configurations in the random sampler as $M=N$. For the channel model, we have used $N_p=3$ paths and the ${\rm SNR}$ was set to $5$dB (this values includes ${\rm PL}_1$). The Normalized Mean Squared Error performance of the estimated beamspace matrix, defined as $\text{NMSE} \triangleq \Vert \mathbf{z}_1 - \mathbf{\hat{z}}_1 \Vert / \Vert \mathbf{z}_1 \Vert$ with $\mathbf{\hat{z}}_1$ denoting the estimation for $\mathbf{z}_1$, is plotted in the figure as a function of the number of training symbols $T$. We have compared the proposed technique with the Orthogonal Matching Pursuit (OMP) algorithm for Multiple Measurement Vectors (MMV) \cite{Alkhateeb_JSTSP_2014}, which exploits the common sparsity pattern between consecutive training slots. The Least Squares (LS) channel estimator is also simulated. As it is evident in Fig$.$~\ref{fig:nmseVSframes_ris}, the proposed technique yields the best estimation performance with fewer training symbols. It is also shown that for achieving a target NMSE value, a larger number of training symbols $T$ is required when $N$ increases.

We finally include an indicative result, due to space limitations, for the achievable end-to-end rate of a RIS-assisted communication system with $N=16$, where each RIS unit element has the phase resolution $b=1$. For the rate computation, we have used the exhaustive search approach of Sec$.$~\ref{sec:tuning} assuming ${\rm SNR}=5$dB for both estimations of the involved channels, as well as for data communication in both links. For perfect channel knowledge and $M=2^{16}$ RIS configurations, the rate is $8.5$bps/Hz, while for channel estimation with $T=200$, the rate is $7.6$bps/Hz, i.e., around $10\%$ lower. 

\bibliographystyle{IEEEtran}
\bibliography{IEEEabrv,references}

\end{document}